\documentclass[preprint2]{aastex}
\usepackage{rotating}
\usepackage{graphicx}
\usepackage{grffile}

\slugcomment{Intended for ApJ, February 2010}
\shortauthors{Crepp et al. 2010}

\begin{document}
\setcounter{secnumdepth}{2}
\title{On-Sky Demonstration of a Linear Band-limited Mask with Application to Visual Binary Stars}

\author{J. Crepp\altaffilmark{1,2}, E. Serabyn\altaffilmark{3},
J. Carson\altaffilmark{3,4,5}, J. Ge\altaffilmark{1}, I. Kravchenko\altaffilmark{1,6}}
\email{jcrepp@astro.caltech.edu}

\altaffiltext{1}{University of Florida, 211 Bryant Space Science Center, Gainesville, FL 32611}
\altaffiltext{2}{California Institute of Technology, 1216 E. California Blvd., Pasadena, CA 91125} 
\altaffiltext{3}{Jet Propulsion Laboratory, 4800 Oak Grove Drive, Pasadena, CA 91109}
\altaffiltext{4}{Max Planck Institute for Astronomy, Heidelberg, Germany} 
\altaffiltext{5}{College of Charleston, 66 George Street, Charleston, SC 29424}
\altaffiltext{6}{Oak Ridge National Laboratory, Oak Ridge, TN 37831}

\begin{abstract}
We have designed and built the first band-limited coronagraphic mask used for ground-based high-contrast imaging observations. The mask resides in the focal plane of the near-infrared camera PHARO at the Palomar Hale telescope and receives a well-corrected beam from an extreme adaptive optics system. Its performance on-sky with single stars is comparable to current state-of-the-art instruments: contrast levels of $\sim10^{-5}$ or better at 0.8" in $K_s$ after post-processing, depending on how well non-common-path errors are calibrated. However, given the mask's linear geometry, we are able to conduct additional unique science observations. Since the mask does not suffer from pointing errors down its long axis, it can suppress the light from two different stars simultaneously, such as the individual components of a spatially resolved binary star system, and search for faint tertiary companions. In this paper, we present the design of the mask, the science motivation for targeting binary stars, and our preliminary results, including the detection of a candidate M-dwarf tertiary companion orbiting the visual binary star HIP 48337, which we are continuing to monitor with astrometry to determine its association.
\end{abstract}

\section{Introduction}
The band-limited mask (BLM) is an occulter located in the image
plane of the Lyot coronagraph. It suppresses telescope diffraction
by manipulating the electric-field amplitude of incident starlight
\citep{kuchner_traub_02}. Compared to other coronagraphic designs,
the BLM makes efficient use of photons \citep{guyon_06}, is robust
to low-order optical aberrations \citep{kcg_05,sg_05,crepp_06} and
finite stellar size (Crepp et al. 2009), and has generated the
deepest contrast in the lab to date \citep{trauger_traub_07}. It is
a viable option for directly detecting terrestrial exoplanets from
space \citep{levine_09} and is currently slated for use with the James Webb Space Telescope
NIRCAM \citep{krist_09}.

Although the BLM shows great promise for conducting large dynamic-range observations, one has never been tested on an astrophysical source. Moreover, the most common type of BLM used for numerical simulations and lab experiments has a linear structure -- as opposed to an azimuthally symmetric (radial) profile -- and can be used to search for tertiary companions by blocking the light from two stars at the same time. This simple geometric feature provides access to an important and as yet unexplored observational parameter space: visual binary stars as high-contrast imaging targets. 

To justify use of a BLM (of any shape) from the ground, strehl ratios in excess of $\approx0.88\:S_{qs}$ are required, where `qs' stands for quasi-static and $S_{qs}<1$ is the Strehl ratio provided by the optical system in the absence of atmospheric turbulence \citep{crepp_07}. This level of correction corresponds to the `extreme' adaptive optics (AO) regime at near-infrared wavelengths. We have access to an AO system and a 1.6m diameter\footnote{Telescope diameter as projected onto the primary mirror.} unobscured and well-corrected off-axis subaperture at Palomar that generates Strehl ratios as high as 96\% in the $K_s$-band under good seeing conditions \citep{serabyn_07}. Upon satisfying the aforementioned criterion, we have built a BLM for on-sky tests and installed it in the Palomar High Angular Resolution Observer (PHARO, \cite{hayward_01}) camera on the Hale telescope. The aim of this paper is to demonstrate the general utility of the BLM and to highlight the importance of studying the immediate vicinity of binaries.


\section{Scientific Motivation}\label{sec:science}
Binaries are a natural result of the star formation process and
constitute $\approx 50\%$ of all stellar systems
\citep{duquennoy_mayor_91}. To date, however, they have been neglected by 
most high-contrast imaging surveys (including both visual and unresolved binaries). In 
the case of visual binaries, this observational bias is due to an
inability to suppress the light from both stars simultaneously.
Depending on the separation and brightness ratio, the presence of a nearby 
second star will add pointing errors or overwhelm a portion of the
final image. Spatially resolving the pair results in significant
contrast degradation and reduces the prospects for discovering faint
tertiary companions.

A linear BLM has an intensity transmission that depends on only one
Cartesian coordinate and can circumvent this problem. Since there
are no pointing penalties down its long axis, simultaneous alignment
to both stellar components can provide high-contrast images of the
circumbinary environment.\footnote{Other coronagraphs are also
capable of targeting visual binaries, but are most often optimized
for single stars.} We now discuss two specific science applications, noting
that unresolved binaries are likewise promising targets.

\subsection{Triple Star Systems}
\label{sec:MSC}
Observational multiplicity studies can provide
important constraints on star formation theories. Consequently, much
work has gone into monitoring the mass ratios and separations of
multiple-star systems as a function of age -- from proto-stars, to
clusters of various size, to mature stars in the field
\citep{mckee_ostriker_07}. From this we have learned that stellar
cores fragment into multiple objects, usually $2-3$
\citep{goodwin_05}, and that dynamical interactions and other
environmental effects \citep{kraus_07} are largely responsible for
determining their final architecture.

Numerical simulations are now able to reproduce many of the bulk
statistical properties observed in young stellar systems
\citep{bate_09}. However, such models still require fine tuning as
they do not yet include all of the relevant physics. Triple stars
offer unique testing grounds for studying the more subtle issues
that binaries alone cannot address, such as angular momentum
orientation, complex accretion and disk truncation, migration, and
the Kozai mechanism \citep{kozai_62}. For example, the existence of a tertiary
companion helps constrain the orbital history of the binary.
Preliminary analyses comparing the properties of triple stars to
binaries and higher-order multiples have helped to shape our
understanding of their origin, their dynamical decay into binary and
single stars, and the physical effects that govern their evolution
\citep{tokovinin_08}. Such efforts are, however, currently limited
by observational biases inherent to the Multiple Star Catalog
(MSC\footnote{http://www.ctio.noao.edu/$\sim$atokovin/stars/},
\cite{tokovinin_97}), which lists all known triples and higher-order
multiples. The MSC is incomplete in the solar neighborhood, missing
$\approx90\%$ of triple systems (several hundred) within 50 pc
\citep{tokovinin_04}.

Efficient pathways towards finding new triple star systems, and thus
improving our knowledge through statistics, include radial velocity
measurements of individual binary components and imaging
observations of the regions surrounding binaries. A coronagraph that
accommodates visual binaries can fulfill this latter role and
uncover low-mass tertiaries that standard AO observations may not 
be capable of detecting.

\subsection{Circumbinary Exoplanets}
\label{sec:exoplanets}
Binary stars must be considered when assessing the statistical properties of exoplanets. In addition to their obvious demographical significance, the binary planet population also has theoretical implications for our understanding of how planets form and evolve. The presence of a second star changes the properties of a protoplanetary disk \citep{artymowicz_94} and, depending on its separation and relative inclination, may either promote or inhibit the growth of dust grains, planetesimals, and embryos \citep{payne_09,xie_09}. Models must be able to account for this extra source of radiation and gravity. For example, the efficiency of the core-accretion process is sensitive to the location of the snow-line \citep{ida_lin_04,kennedy_08}, which is clearly more complicated in a double-star system. Another leading theory, gravitational instability (Boss 1998), predicts that gas-giant planet formation may be enhanced by a nearby star: injecting turbulence into the disk increases the likelihood for fragmentation. Theoretical models would clearly benefit from observational constraints in addition to those imposed by single stars alone.

So far, the search for planets in binaries has focused primarily on imaging systems with wide ($>$ 100 AU) separations \citep{bonavita_07,mugrauer_07}. However, in recognition of the fact that more closely-spaced binaries can provide unique theoretical testing grounds, radial velocity surveys have begun to target systems with much smaller separations \citep{hatzes_03,eggenberger_07}. These efforts have already revealed an important result: short-period planets in S-type (circumstellar) orbits tend to be more massive than their single star counterparts \citep{eggenberger_09,duchene_10}. An imaging search for planets in P-type (circumbinary) orbits would complement this work and provide further constraints on formation theories (see also \cite{muterspaugh_07}). Nevertheless, very few teams have established such programs and those that have use indirect techniques that are relatively insensitive to distant companions. As a result, only a handful of circumbinary planets are currently known, anywhere from 2-4 depending on where the dividing line between brown dwarfs and planets is drawn \citep{lee_09,backer_93,correia_05}.

Based on the limited sample size of current surveys (e.g., \cite{konacki_09} have observed 10 systems using radial velocity techniques with a baseline of 5 years at 2 m/s precision), these likely account for only a small fraction of the population. Other data sets suggest that circumbinary planets may be abundant: (i) circumbinary disks have been detected around pre-main-sequence systems, such as CoKu Tau 4 \citep{ireland_08}, KH 15D \citep{herbst_08}, AK Sco \citep{andersen_89}, GW Ori \citep{mathieu_91}, DQ Tau \citep{mathieu_97,boden_09}, and GG Tau \citep{dutrey_04}; (ii) moderately-separated binaries do not appear to adversely affect the first steps of planet formation -- the growth and crystallization of sub-micron dust grains \citep{pascucci_08}; (iii) a recent Spitzer study indicates that $\sim60\%$ of binaries with separations $<3$ AU display excess thermal emission from dusty circumbinary debris produced by planet-induced collisional cascades \citep{trilling_07}; and (iv) simulations indicate that tertiary companions may remain stable in such configurations for billions of years, over a wide range of mass ratios and eccentricities \citep{holman_wiegert_99}.

Finally, visual binaries may also be uniquely useful targets because they
can be selected to have near face-on orbits. Assuming circumbinary
planets form in roughly the same orbital plane as their stellar
hosts, a survey that targets binaries with opportunistically
selected orientations will have a higher detection efficiency
compared to those that observe systems with unknown inclinations.


\section{Mask Design \& Fabrication}
\label{sec:design}

\begin{figure}[!ht]
\begin{center}
\includegraphics[height=1.6in]{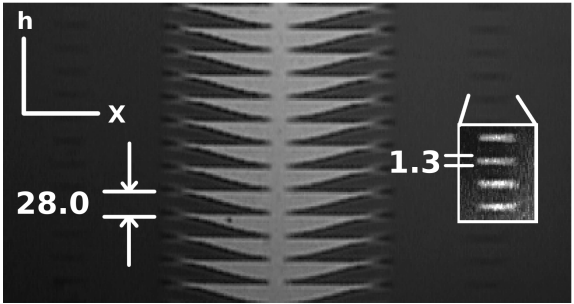}
\caption{Microscope image of the BLM prior to cleaning. Each stripe is $28 \: \mu$m tall, slightly smaller than a resolution element in $K_s$. `Ringing' features from the $\mbox{sinc}^2$ function (Equ.~\ref{eqn:mask_pharo}), which are responsible for eliminating diffraction, can be seen on either side of the main occulting region.}\label{fig:blm}
\end{center}
\end{figure}


The bandpass and focal ratio, $f$, constrain the mask design. Fast optical systems can force the size of features to be prohibitively small, whereas broad wavelength coverage affects the inner-working angle (IWA) and Lyot stop throughput. PHARO accepts an $f=15.64$ beam from the Palomar AO system \citep{troy_00}. The $K_s$ filter served as our primary science channel and defined the desired operating range for observations, $\lambda=1.99-2.31 \; \mu$m. We decided to build a smooth, binary, fourth-order, linear BLM based on these parameters, our available tools, experience \citep{crepp_06}, and the aforementioned science motivation. An eighth-order mask \citep{kcg_05}, which has less Lyot stop throughput than a fourth-order mask, could not be justified, given the reduction in flux due to the 1.6m diameter subaperture effective size.


A binary linear BLM consists of a series of small stripes (Fig.\ref{fig:blm}). The key to its operation is that each stripe is smaller than the resolution of the optical system, $f \; \lambda_{\mbox{min}}$, hence allowing it to diffract light like a graded mask \citep{kuchner_spergel_03}. To prevent diffracted light leakage from the edges of the filter transmission profiles, we chose conservative values for the minimum and maximum design wavelengths: $\lambda_{\mbox{min}}=1.79 \;\mu$m and $\lambda_{\mbox{max}}=2.39 \;\mu$m. The mask IWA, defined here as the location where the intensity transmission reaches 0.5, is $2.678 \;\lambda_{\mbox{max}} / D_{\mbox{tel}} = 880 \;$mas. The height,
$h(x)$, of a single stripe of the BLM is given by:
\begin{eqnarray}
h(x)&=& f \:
\lambda_{\mbox{min}}\left[1-\mbox{sinc}^2\left(\frac{\pi\epsilon
x}{2 \: f \: \lambda_{\mbox{max}}} \right) \right]
\label{eqn:mask_pharo}
\\  &=& 28 \left[1-\mbox{sinc}^2\left(\frac{0.0179857 \:
x}{\mu\mbox{m}}\right) \right] \; \mu\mbox{m}, \nonumber
\end{eqnarray}
where $\epsilon=0.428$ and $x$ has units of $\mu$m. This profile was repeated vertically 256 times. 


The BLM was fabricated using electron-beam lithography at the University of Florida nanofabrication facility. Aluminum served as the opaque material deposit. Lab measurements placed an upper limit on the intensity transmission directly through the $\sim$200 nm layer at $1\times10^{-7}$. The minimum feature size was set equal to the layer thickness. The mask was placed on a single 0.7 mm thick Boro-Aluminosilicate glass substrate, Corning model 1737, with a K-band transmission quoted at 92\%. No anti-reflection coating was applied.

Figure~\ref{fig:blm} shows images of the mask before it was shipped to the Jet Propulsion Laboratory (JPL), where it was cut from the substrate with a dicing saw to physical dimensions of 0.60 x 0.30 inches and then cleaned in an ultrasonic bath of acetone to remove a protective layer of photoresist that was applied prior to shipping. We ran simulations to ensure that truncation at the mask edges would not diffract starlight into the search area at contrast levels above $10^{-6}$.
An Aluminum holder was built to fasten the BLM to the PHARO slit wheel. 



\section{On-Sky Demonstration}
\label{sec:onsky}
\subsection{Single Stars: $\epsilon$ Eridani}
The star $\epsilon$ Eridani is an attractive target to demonstrate
new coronagraphic technologies (e.g.,
\cite{macintosh_03,proffitt_04,itoh_06,janson_07,janson_08,heinze_08}). It is nearby
($d=3.2$ pcs), bright (V$\sim3.7$), relatively young
(age$\approx730$ Myrs \citep{song_00}), known to host a lumpy debris
disk \citep{greaves_98}, and shows radial velocity and astrometric
signatures consistent with that of at least one substellar companion
with a long period \citep{hatzes_00,benedict_06}.

We observed $\epsilon$ Eridani (K2V) on 11-07-2008 UT to test the BLM's
on-sky performance. The seeing was $\approx0.5"$ in the
$K_s$ band and the AO system was running at 1 kHz,
providing an on-axis Strehl ratio of $\sim 90\%$ (see Serabyn et al.
2007 for a discussion of the well-corrected subaperture). We
acquired 30 occulted images of $\epsilon$ Eridani with an
integration time of 14.16s each and 5 unocculted images with an
integration time of 4.25s each, using a well-characterized
neutral-density filter to calibrate the photometry. We also acquired
15 occulted images of a nearby calibrator star, $\delta$ Eridani, to
suppress quasi-static speckles via PSF subtraction. Only several
minutes elapsed on average between acquisition of science images and
those of the calibrator to minimize evolution of the speckle
pattern. This technique has been successfully employed before at
Palomar by \cite{mawet_09} and Serabyn et al. (2010) who used a phase-mask 
coronagraph to characterize the debris disk around HD 32297 and reconfirm 
detection of the HR 8799 planets in the near-infrared. 


We processed the data with an in-house reduction code written in
Matlab. The procedure utilizes standard flat-fielding, bad pixel
correction, and sky background subtraction techniques. After
applying a quadratic high-pass Fourier filter to each occulted
image, to remove the smooth component of the speckle halo, we used
the `locally optimized combination of images' (LOCI) algorithm \citep{lafreniere_07} 
with a box size of $3\times3$ PSF FWHM's to perform PSF subtraction.

\begin{figure*}[!t]
\begin{center}
\includegraphics[height=2.12in]{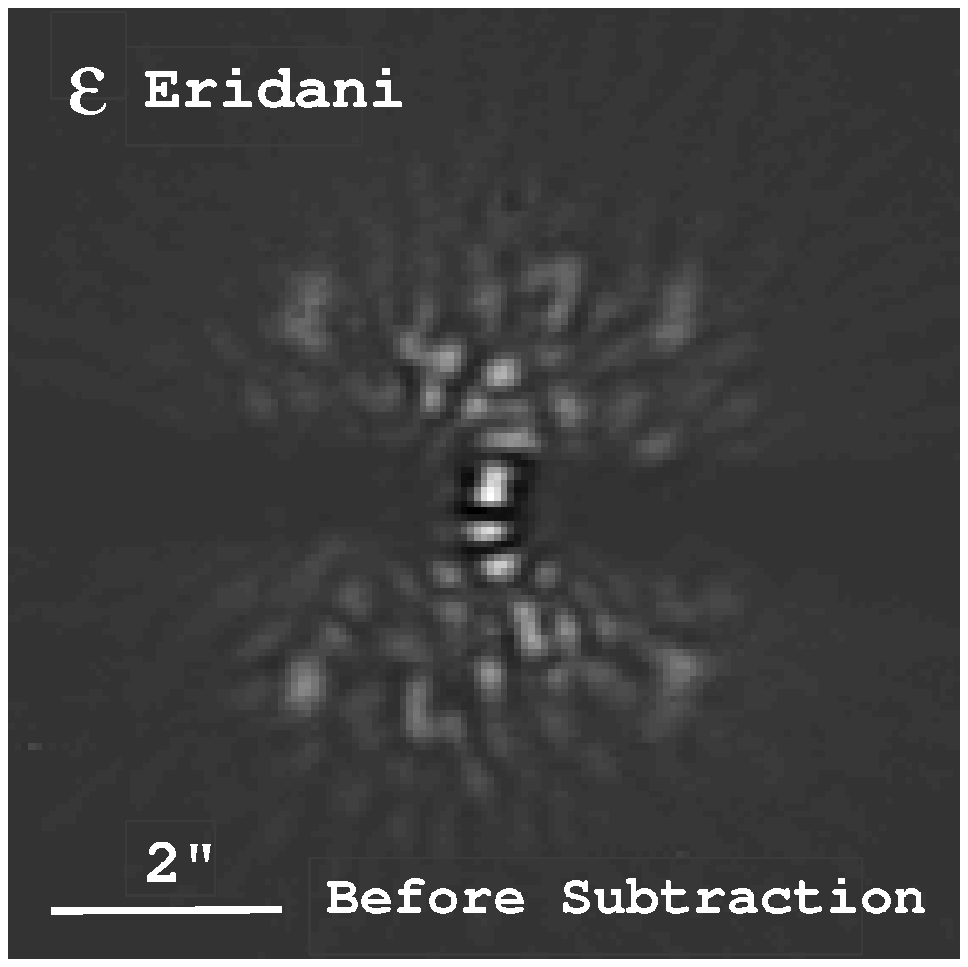}
\includegraphics[height=2.12in]{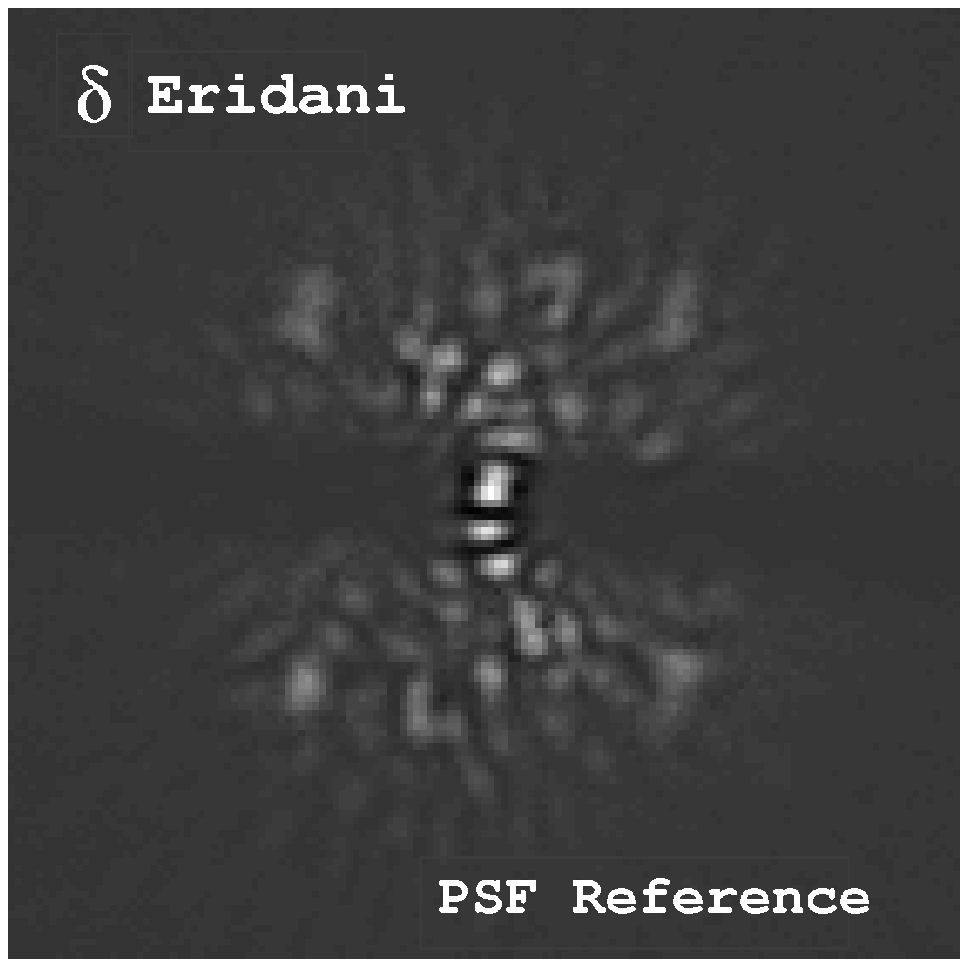}
\includegraphics[height=2.12in]{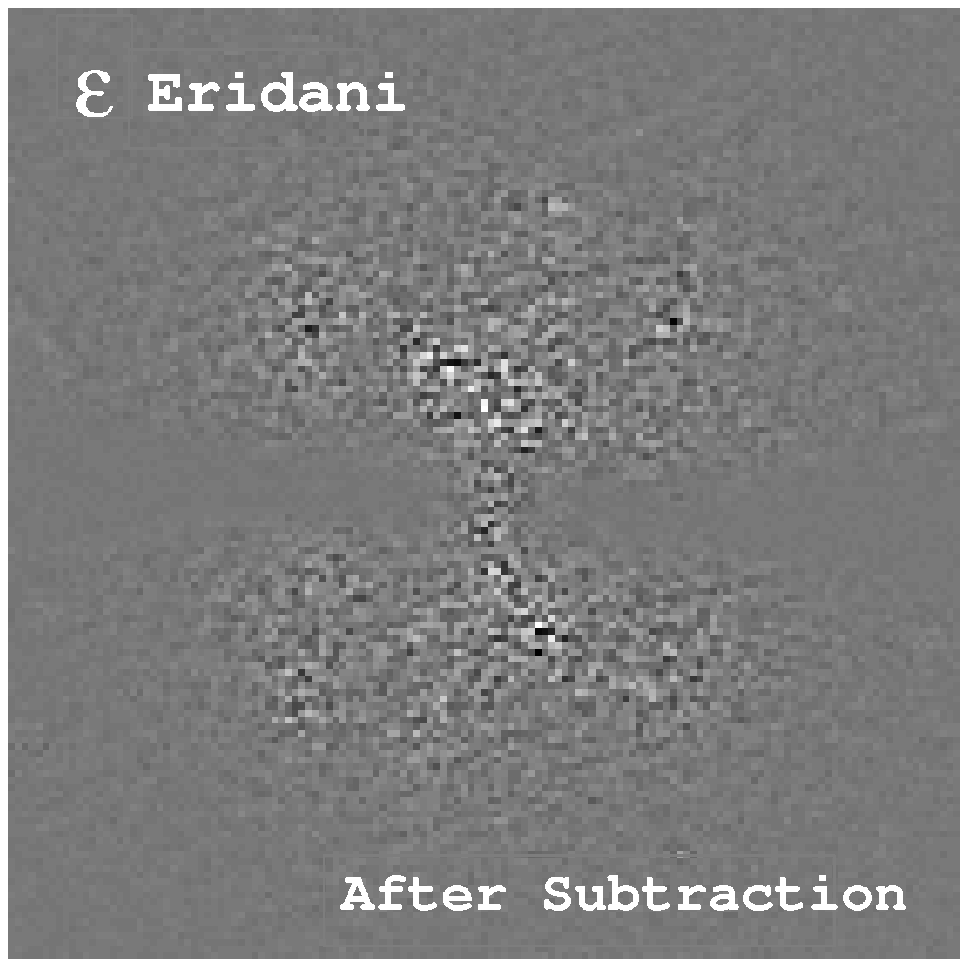}
\caption{Fully processed $K_s$ images of the star $\epsilon$ Eridani and a nearby calibrator, $\delta$ Eridani, using the BLM. (left) High-pass Fourier filtered image of the target star prior to PSF subtraction. (middle) PSF calibrator constructed from 15 separate images with the LOCI algorithm. The speckle patterns are highly correlated. (right) Differenced image.}
\label{fig:eps_eri}
\end{center}
\end{figure*}

Fig.~\ref{fig:eps_eri} shows images of the target star, calibrator,
and subtraction residuals. The BLM clearly removes the vast majority
of diffracted starlight. Only a hint of the first two Airy rings
remain in the pre-subtraction images; these residuals are due to the
fact that we purposefully use a slightly over-sized Lyot stop to
increase the throughput of companions and improve spatial
resolution. We find that a smaller Lyot stop enables the BLM to
remove all on-axis diffracted light in the lab. On-sky Airy
rings are extremely stable and subtract-out better than speckles -- when using the 
well-corrected subaperture, though minimizing their initial intensity is likewise important.

\begin{figure*}[!ht]
\begin{center}
\includegraphics[height=3.8in]{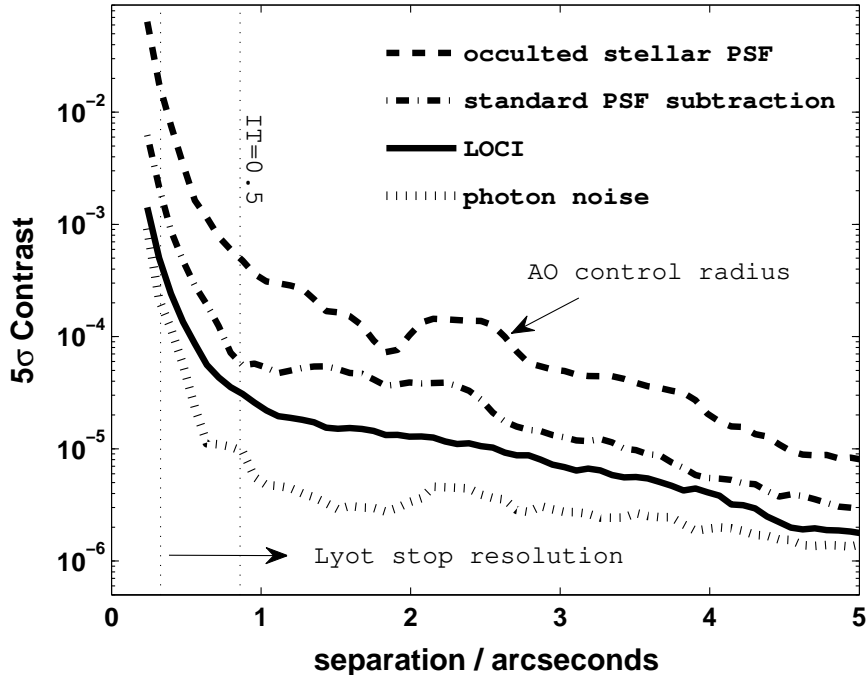}
\caption{Contrast levels ($5\sigma$) before and after PSF subtraction for the star $\epsilon$ Eridani in $K_s$. The BLM working in concert with the well-corrected subaperture and LOCI algorithm allows for the unambiguous detection of companions $\sim 10^5$ times fainter than the star in the primary search region. The standard PSF subtraction and LOCI curves each include the contributions from speckle and photon noise. The distance where the BLM intensity transmission (IT) reaches 0.5 is indicated by a vertical line.}\label{fig:contrast}
\end{center}
\end{figure*}

To calculate our sensitivity in terms of contrast, we added the photon-noise and speckle noise in quadrature. Fig.~\ref{fig:contrast} shows results for a 2 $\lambda_{\mbox{max}}/D_{\mbox{tel}}$ wide region running perpendicular to the mask axis, where the contrast above and below the BLM has been averaged. The LOCI algorithm provides a factor of $\sim 2-3$ improvement over standard PSF subtraction. We were able to reach $5\sigma$ contrast levels of $3.5\times10^{-5}$ at 0.8", $1.3\times10^{-5}$ at 2.0", and $4.0\times10^{-6}$ at 4.0", which correspond to masses of $\approx24.0M_J$, $19.2M_J$, and $14.7M_J$ respectively at an age of 730 Myrs \citep{baraffe_et_al_03}. Though no companions were detected,\footnote{The inverse operation, using $\epsilon$ Eridani as a calibrator for $\delta$ Eridani, also did not reveal any candidates. } these observations represent the most sensitive $K_s$ high-contrast images of $\epsilon$ Eridani inside of 5" to date, even though the aperture used is much smaller than in previous studies (see aforementioned references). 


More images would allow us to improve PSF subtraction by
constructing a better calibrator reference with the LOCI algorithm
and approach the limit set by the photon-noise of the speckles. To
decrease this floor further, the non-common-path errors between the
AO system and PHARO must be reduced. Precision calibration of
non-common-path errors would also improve the speckle symmetry and
enable the target star to potentially serve as its own calibrator.
Indeed, some symmetry is noticeable in the images in
Fig.~\ref{fig:eps_eri}. At the time of our observations, the
deformable mirror actuator offsets were tuned manually with the
first 10 Zernike polynomials by monitoring the PHARO PSF, resulting
in a wavefront error of $\sim 110$ nm rms. Work is currently
underway to reduce the initial intensity of static speckles with
robust diversity-based phase retrieval techniques, such as the
\cite{gerchberg_saxton_72} algorithm \citep{sidd_08}. Initial results show promise
to improve contrast levels by as much as a factor of $\approx(110\:
\mbox{nm} \:/ \: 30 \:\mbox{nm})^2=13$ with the current 17x17
element deformable mirror (R. Burruss, private communication).



\subsection{Visual Binaries: HIP 48337}
We observed HIP 48337 on 11-10-2008 UT as part of a mini-pilot survey to search for low-mass tertiary companions using the linear BLM. Listed as a binary in SIMBAD and NStED, HIP 48337 has a `reliably' well-determined orbit (grade=3/5) from speckle interferometry measurements, having coverage over more than half of a period according to the USNO 6th Catalog of Visual Binary Orbits \citep{mason_02}. Its components have similar near-infrared magnitudes (Table~\ref{tab:hip48337}) and were separated by 0.35" at the time of our first epoch observations.

\begin{figure}[!ht]
\begin{center}
\includegraphics[height=3.1in]{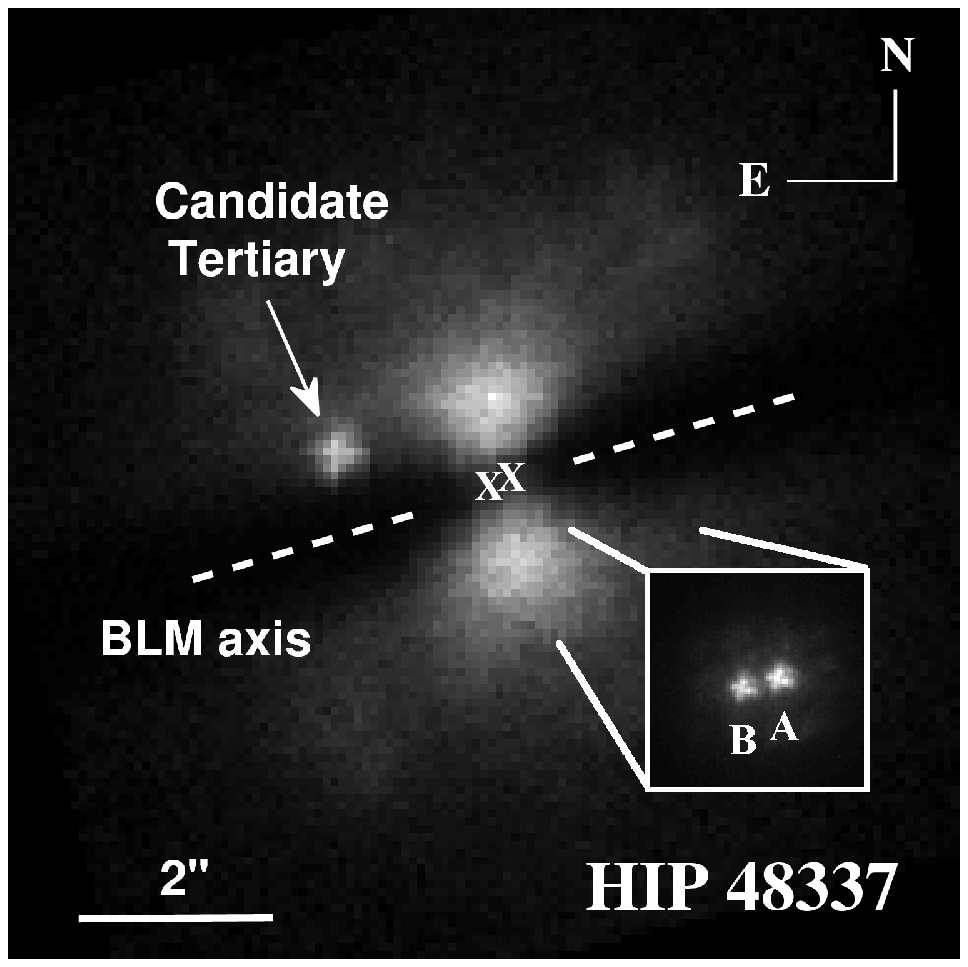}
\caption{Discovery image of the HIP 48337 C candidate companion taken at Palomar on 11-10-2008 UT in $K_s$. The primary components, HIP 48337 A,B, are spatially resolved, as shown in the unocculted subimage, and simultaneously attenuated by the linear BLM. The locations of HIP 48337 A,B in the occulted image are each denoted by an `x'. The BLM axis is marked by a dashed line. }
\label{fig:hip48337_image}
\end{center}
\end{figure}

Upon aligning the linear BLM to block each source, we noticed a
faint candidate tertiary companion separated by 1.95" from HIP 48337 A. Fig.~\ref{fig:hip48337_image} shows the original discovery image. The candidate is visible in raw data and comparable in brightness to the AO residuals, which are dominated by temporal
errors since HIP 48337 is relatively faint for AO locking (combined apparent magnitude 
of V=7.89) with the well-corrected subaperture. Although well-separated 
from the two primary stars, this detection demonstrates the utility of a linear 
coronagraphic mask. We could have unambiguously identified the candidate in less than 10 minutes at the same position angle for separations as small as 0.9". If it is gravitationally bound, then HIP 48337 is one of the many nearby triples missing from the MSC ($\S$\ref{sec:MSC}).


We measured the relative brightness of each HIP 48337 component in the J, H, K bands
using standard aperture photometry, taking into account minor contamination from the
other components. These numbers were then converted to apparent magnitudes using 
the seeing-limited (unresolved) 2MASS measurements of the system combined brightness 
through the equations:
\begin{eqnarray}
m_A&=&m_{AB}+2.5 \;  \mbox{log}(1+f_B/f_A) \\
m_B&=&m_{AB}+2.5 \; \mbox{log}(1+f_A/f_B) 
\end{eqnarray} 
where $m_A<m_B$ are the apparent magnitudes of HIP 48337 A, B respectively, $m_{AB}$
is the combined apparent magnitude from 2MASS, and $f_A>f_B$ are the individual component fluxes in arbitrary units. The apparent magnitude of the HIP 48337 C candidate, which was too faint to be detected by 2MASS, was then calculated from $f_C$. The results are 
shown in Table~\ref{tab:hip48337}.

Converting to absolute magnitudes and comparing the results with \cite{girardi_02} isochrones suggests the A, B, C components have spectral types of A8-F8, F2-G0, and M0-M4 respectively, assuming they lie on the main-sequence. These inferences are limited by the uncertainty in distance to HIP 48337 ($d=133.0 \pm 41.9$ pcs). We find that results based on colors alone, which are independent of distance, have similar uncertainties (e.g., comparing to \cite{ducati_01}). More detailed characterization using spatially-resolved spectra and precision astrometry from the Project 1640 instrument (see \cite{zimmerman_10,hinkley_10}) will be reported in a subsequent paper.  

\begin{table}[!h]
\centerline{
\begin{tabular}{cccc}
HIP 48337       &                   A                  &                 B                    &                  C                       \\
\hline
$V$                   &    $8.54\pm0.01$          &  $8.75 \pm 0.02$        &        ---                       \\   
$J$                    &    $7.44 \pm 0.06$        &   $7.73 \pm 0.07$      &    $12.70 \pm 0.16$    \\
$H$                   &   $7.23 \pm 0.06$         &    $7.54 \pm 0.07$      &    $12.15 \pm 0.15$    \\
$K$                   &   $7.18 \pm 0.06$         &    $7.51 \pm 0.08$       &     $11.82 \pm 0.16$    \\
$M_V$             & $2.92 \pm 0.83$           &    $3.13 \pm 0.84$         &      ---                           \\
$M_J$              &  $1.82 \pm 0.88$          &     $2.11 \pm 0.89$       &    $7.08 \pm 0.98$       \\
$M_H$             &  $1.61 \pm 0.88$          &     $1.92 \pm 0.89$      &  $6.53 \pm 0.98$          \\
$M_K$             &  $1.56 \pm 0.88$          &     $1.89 \pm 0.89$      &  $6.20 \pm 0.98$           \\
$\rho$ (mas)    &     ---                                &    $354 \pm 6$            &       $1952 \pm 13$      \\
$\theta$ (deg)  &     ---                                &   $101.6 \pm 0.2$      &      $86.0 \pm 0.2$        \\
\hline
\end{tabular}}
\caption{Physical parameters of HIP 48337. Visual magnitudes are from Hipparcos. Absolute magnitudes for the HIP 48337 C candidate are calculated for an assumed distance of 133 pcs. Angular separations and position angles correspond to observations conducted on 03-21-2009 UT.}
\label{tab:hip48337}
\end{table}

\section{Summary}
We have designed and built the first band-limited mask (BLM) used for ground-based high-contrast imaging observations. The BLM is installed in the PHARO camera at Palomar and is capable of generating $5\sigma$ contrast levels of order $10^{-5}$ at subarcsecond separations in the $K_s$ band when operating in tandem with an extreme AO system and performing PSF reference subtraction. We have demonstrated this technology on the star $\epsilon$ Eridani, achieving a mass sensitivity of $\approx24.0M_J$ at 0.8" and $\approx14.7M_J$ at 4.0". These observations are the most sensitive in $K_s$ yet reported for the star inside of 5" to date, even though the aperture used is much smaller than in previous studies. Further improvements to calibration of non-common-path errors between the AO system and PHARO will provide even deeper contrast. 

We have outlined two science cases that motivate binary star high-contrast imaging. The search for tertiary companions in P-type orbits can significantly improve our understanding of: (i) star formation, by improving the statistics of the Multiple-Star Catalog \citep{tokovinin_97,tokovinin_04}, which is currently incomplete beyond $\sim10$ pcs, and (ii) planet formation, by complementing the efforts of radial velocity teams searching for planets in S-type orbits. We have already detected a candidate M-dwarf tertiary orbiting the visual binary HIP 48337, showing that a linear BLM is a viable tool for exploring this observational parameter space. It is possible to speculate that the companion, if gravitationally bound, may be exciting the eccentricity of HIP 48337 A,B via the Kozai mechanism: their eccentricity is $\sim 0.93$ according to the USNO WDS catalog. 

We thank Karl Stapelfeldt for support at the JPL microdevices lab to cut and clean the BLM prior to installation and Dimitri Mawet for helpful discussions at the observatory. J. Carson acknowledges support from the NASA postdoctoral program. This work was funded in part by the UCF-UF-SRI program and NASA grant NNG06GC49G.

\begin{footnotesize}
\bibliographystyle{jtb}
\bibliography{ms.bib}
\end{footnotesize}

\end{document}